\documentclass[structabstract]{aa}  
\usepackage{graphicx}
\usepackage{txfonts}
\usepackage{natbib}
\def\beq{\begin{equation}}
\def\eeq{\end{equation}}

\def\msol{\,\mathrm{M}_\odot}
\def\M{M_\bullet}

\def\wbh{W_\mathrm{BH}}

\begin{document}

\defcitealias{mio07}{M07}

   \title{A mass estimate of an intermediate-mass black hole in
$\omega$ Centauri}

   \subtitle{}

   \author{P. Miocchi
          }

   \institute{
INAF, Osservatorio Astronomico di Bologna, and
Dipartimento di Astronomia, Universit\'a di Bologna, \\
via Ranzani, 1, Bologna I-40127, Italy.\\
              \email{paolo.miocchi@unibo.it}
             }

   \date{}

 
  \abstract
   {The problem of the existence of intermediate-mass black holes (IMBHs)
at the centre of globular clusters is a hot and controversial topic in current
astrophysical research with important implications in stellar and galaxy
formation.}
   {
In this paper, we aim at giving further support to the presence
of an IMBH in $\omega$ Centauri and at providing an independent estimate of
its mass. }
   {
We employed a self-consistent spherical model
with anisotropic velocity distribution. It consists in a generalisation
of the King model by inclunding the Bahcall-Wolf distribution
function in the IMBH vicinity.}
   {
By the parametric fitting of the model to recent HST/ACS data for
the surface brightness profile, we found an IMBH to cluster total mass ratio of
$\M/M=5.8_{-1.2}^{+0.9} \times 10^{-3}$. It is also found that the model yields
a fit of the line-of-sight velocity dispersion profile that is better without
mass segregation than in the segregated case.
This confirms the current thought of a non-relaxed status for this peculiar cluster.
The best fit model to the kinematic data leads, moreover, to a cluster total mass
estimate of $M=(3.1 \pm 0.3) \times 10^6 \msol$, thus giving an IMBH mass in the range
$1.3\times 10^4 < \M < 2.3 \times 10^4 \msol$ (at $1\sigma$ confidence level).
A slight degree of radial velocity
anisotropy in the outer region ($r \ga 12\arcmin$) is required to
match the outer surface brightness profile.}
   {}


   \maketitle
%

\section{Introduction}

Intermediate-mass black holes (IMBH) still belong to the class of
`exotic' objects in the current astrophysical belief.
With masses between $\M\sim 100$--$10^4 \msol$, they
would represent the minor mass counterpart
of super-massive black holes -- whose existence is established with
much more robustness -- but still more massive than stellar black holes.
One of the places where they should
more likely be located is among the densest stellar environments
in the Universe, i.e. at the globular clusters (GCs) centre.
However, so far, the most direct observable signature of their
existence, namely the emission in the radio
and X-ray bands (mainly from Bondi-Hoyle accretion of intracluster
gas), is not yet really clear and conclusive 
(see, e.g., \citealt{liu08}; \citealt{zepf08}; \citealt{irwin09}; \citealt{stro09};
and \citealt{miller04} for a general review).

To date, only the GC G1 (in M31) exhibits a detected source, seen in both radio 
\citep[with an $8.4$ GHz power of $2\times 10^{15}$ W Hz$^{-1}$, see][]{ulvestad} and X-ray
\citep[with a $2\times 10^{36}$ erg s$^{-1}$ luminosity at
$0.2$--$10$ keV, see][]{pooley06} bands. The observed fluxes, as well as their ratio,
are compatible with the claimed presence of a $\sim 2\times 10^4\msol$ IMBH \citep{gebhardt} --
although other kinds of sources cannot be completely ruled out \citep[e.g.][]{kong}.
Another extra-galactic hyperluminous
X-ray source ($5$ -- $100 \times 10^{40}$ erg s$^{-1}$ at $0.3$ -- $10$ keV) is
located in the S0-a galaxy ESO243-49
and its features suggest an IMBH emission.
Recently, an unresolved optical counterpart with
brightness comparable to that of a massive GC has been
identified around this source \citep{soria},
though higher resolution observations are needed.

In our Galaxy, the central region of NGC 6388 hosts an unresolved set of X-ray sources,
with a total luminosity of
$2.7\times 10^{33}$ erg s$^{-1}$ \citep{nucita}, implying an accretion
efficiency compatible with the inferred presence of a $\sim 6\times 10^3 \msol$
IMBH \citep{lanzoni}.
On the other hand,
no detectable X-ray sources have been found at the centre of mass of NGC 2808,
leading \citet{servillat09} to state that $\M\la 290 \msol$ in this cluster.

In fact, in most cases only upper limits for IMBHs masses can be deduced from
radio observations \citep[see, e.g.][for NGC 2808 and for a general discussion]
{maccarone08}.
These surprisingly low upper limits lead \citet{maccarone08} to
cast doubts on the fact that the scaling relation
$\M$ -- $\sigma$, where $\sigma$ is the central velocity dispersion
of the host stellar system (with mass $M$ and luminosity $L$), is the same
$\M\sim \sigma^{4.8}$ law that has been clearly noted for
super-massive black holes in galaxies \citep{ferrarese,gebhardt00}.
It should be emphasised, however,
that the upper limits on IMBH masses drawn from X-ray or radio
observations strongly depend on the assumption that the gas distribution
around the compact object is uniform (isotropic accretion).
It is clear that, if this distribution had any amount of clumpiness, those limits
could be largely underestimated.

Nevertheless, the question of the validity of the extrapolation of the
$\M$ -- $\sigma$ scaling relation to IMBHs is still open and
deserves to be discussed briefly here.
In general, this relation can be understood as a consequence of the fundamental
scaling law $\M\propto M$ \citep{magorrian}.
In galaxies, this scaling law and the two relations
$M\sim L^{5/4}$ \citep{faberetal} and
$L\sim \sigma^4$ \citep{faberjackson}, lead just to $\M\sim \sigma^5$.
In globular clusters, on the other hand, the observed trends are
$M\sim L$ and $L\sim \sigma^{5/3}$ \citep{heggiemeylan}, which in fact yield
$\M\sim \sigma^{1.7}$. This implies a generally lower mass ratio between
the IMBH and the host cluster, as noted by \citet{maccarone08}.
A shallow $\M$ -- $\sigma$  relation, namely $\M \sim \sigma^{1.2}$,
was already reported in \citet[][hereafter \citetalias{mio07}]{mio07} based on
parametric IMBH mass estimates (see below).
On the other hand, we must mention a recent study of this specific
topic, in which
the low-mass extrapolation of the galactic $\M\sim \sigma^{4.8}$ relation
seems to fit a sample of 5 reported IMBHs in GCs \citep{safonova}.

In view of all this, the study of possible IMBH fingerprints on
either star-count or surface brightness (SB) profiles
is a detection route that still deserves to be pursued, especially when 
kinematic observations close to the IMBH gravitational
influence region are available.
In this respect, a spherical and self-consistent model
of GCs with a central IMBH at rest was presented in \citetalias{mio07}, both with equal mass stars,
i.e. the single-mass (SM) case, and with a multimass (MM) stellar spectrum
inclunding mass segregation.
This model is an extension of King-Michie models
\citep{michie63,boden63,king66} that is obtained by including
the Bahcall-Wolf stellar distribution function
within the IMBH gravitational influence region.
The latter was shown to solve the Fokker-Planck equation
in the vicinity of a central IMBH that formed long before 
a cluster relaxation time \citep{bw,BT}, and its validity was
subsequently confirmed by accurate numerical simulations
\citep{freitag02,bau04a,preto}.

The typical SB profile that comes out of the model
has, for any reasonable IMBH mass, the appearance of a normal low- or
medium-concentration cluster
($c \la 2$) and shows a steep cusp only in the very inner region (typically
within a tenth of the core radius) delimited by the `cusp radius', $r_\mathrm{cu}$.
In fact, outside $r_\mathrm{cu}$, a shallow power-law behaviour -- with
a logarithmic slope $s\la 0.25$ -- is the most easily
observable fingerprint in the otherwise flat core profile.
Interestingly, this confirmed the finding of other authors
who -- using a completely different approach (accurate $N$-body simulations) -- 
also claim that IMBHs most likely reside in non--core-collapsed clusters
showing just a weak rise of the SB in the core region (\citealt{bau05}; \citealt{trenti07}).
Recently, high-resolution Montecarlo simulations have provided another
independent confirmation of these structural features (\citealt{umbreit09}).
On the other hand,
according to other $N$-body experiments, it is claimed that post--core-collapsed
GCs also exhibit a King-like profile, but with a $s\sim 0.4$ -- $0.7$
steep core behaviour \citep{trenti09};
it must be emphasised, however, that \citetalias{mio07} models yield core
profiles that are always significantly flatter and unable to fit behaviours with
such a high $s$.

The shape of the SB profile given by the \citetalias{mio07} model depends
on 2 dimensionless parameters\footnote{The model can also include velocity anisotropy in
the GC outskirts. In this case
the outer SB profile shape depends on the anisotropy radius, too.}.
For the purposes of this study, we use the IMBH to cluster mass ratio, $\M/M$,
and the dimensionless gravitational potential at the edge of
the IMBH dynamical influence region, $\wbh$.
The latter replaces the usual King model's central dimensionless potential $W_0$,
with the aim of avoiding the singularity at the
cluster centre in the presence of the IMBH \citepalias[see][for further details]{mio07}.
 


In \citetalias{mio07} it was shown that lower and upper limits of $\M$
exist as a function of $c$ and $s$.
This relationship was then applied to investigate the presence of
IMBHs in the set of GCs, whose SB was accurately measured in \citet{noyola06}
using HST/WFPC2 archive images.
Among the six candidate clusters found, NGC 6388 and M54 have subsequently
been the objects of further
and more detailed studies (through parametric fitting
of star-count profiles) that suggest the presence of an IMBH with
mass $\sim 6\times 10^3 \msol$ in the former \citep{lanzoni} and $\sim 10^4$
in the latter
\citep[][in this case kinematic data were also exploited]{ibata}.
On the other hand, the massive cluster $\omega$ Cen was
not checked as a possible candidate, because it was not included 
in the \citet{noyola06} sample, and moreover, 
small slopes in the core region could not be revealed in published
SB profiles \citep[e.g.][]{meylan87,ferraro06}.

Nonetheless, a recent and accurate determination of the $\omega$ Cen centre
and the use of HST/ACS images led \citet{noyola08} to detect
a steeper profile in the core region of this peculiar cluster, thus
suggesting the influence of an IMBH. By fitting the high inner peak of
the line-of-sight velocity dispersion
(LOSVD) found from Gemini GMOS/IFU
integrated light spectroscopy ($=23 \pm 2$ km s$^{-1}$
at an average radius $\sim 1\arcsec.9$) with non-parametric and orbit-based models
with uniform mass-to-light ratio,
\citeauthor{noyola08} estimate a $\sim 4 \times 10^4 \msol$ object residing at the cluster
centre.
However, by solving the spherical and anisotropic Jeans equation on the
\citet{anderson09} projected density and kinematic data,
\citet{anderson09b} find that the presence of an IMBH is possible
only if $\M/M \la 4.3\times 10^{-3}$, which corresponds to about half the mass
predicted by \citeauthor{noyola08}.

In this paper we would like to provide further evidence on the presence of the IMBH
and to give another independent estimate of its mass, by means of a parametric
fitting of both the SB and the LOSVD profiles using the \citetalias{mio07} model.
The results
from the best fit of the SB profile are described in
Sect.~\ref{surf_fit}, while those coming from the LOSVD fitting
are presented in Sect.~\ref{losvd_fit}. Concluding remarks are
reported in Sect. \ref{concl}. 

\section{The fit of the surface brightness}
\label{surf_fit}
To study the $\omega$ Cen SB profile, we considered the HST/ACS measurements
recently made by \citet{noyola08} inside $40\arcsec$ from the cluster centre,
while for outer radii we took the
observations by \citet[][their Table 1]{meylan87}. 

\begin{figure}
\includegraphics[width=9cm]{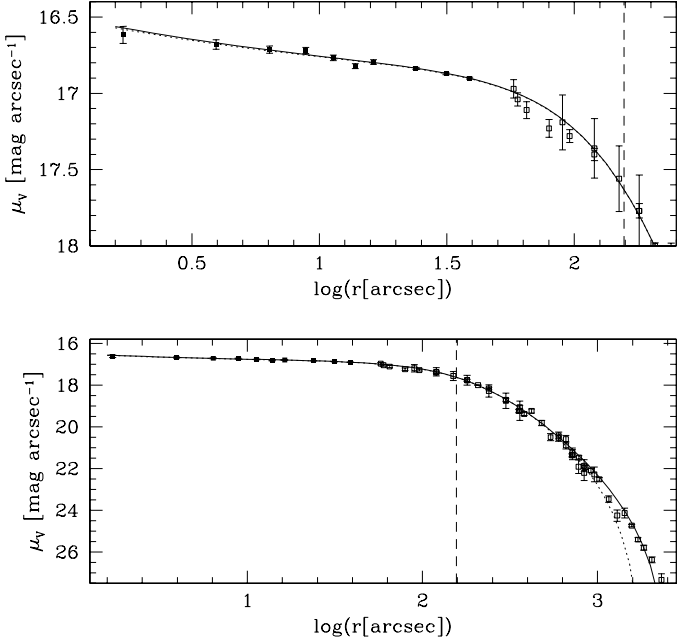}
\caption{Bottom panel: surface brightness profile of $\omega$ Cen and the
SM isotropic (dotted line) and anisotropic (solid line; with $r_\mathrm{a}=4.5r_\mathrm{c}$)
best fit.
In the case of the isotropic model, only data with 
$\log(r)<2.9$ have been considered for the best fit search.
The MM isotropic model yields a best fit profile that is indistinguishable
from that given in the SM anisotropic case.
For $\log(r)<1.6$,
the HST/ACS observations by \citet{noyola08} are used (filled squares), while for $\log(r)>1.6$
data are taken from \citet[][open squares]{meylan87}. The top panel shows an enlarged view of the
central region. The core radius is plotted with a dashed line. 
\label{sb}}
\end{figure}

As is evident from the too low concentration of the dotted profile in Fig.~\ref{sb}
(bottom panel), we notice that the SM isotropic model is unable to 
fit the outermost part ($r \ga 13\arcmin$) of the SB profile, where, however,
the background contamination should be negligible,
for it was shown to only be relevant for $r\ga 33\arcmin$ \citep{leon00}.
Thus, the discrepancy from the prediction of this model should be due
to the intrinsic dynamical state of the cluster outskirts. 
In fact, we find that a good fit of the whole profile can be achieved by including 
either a certain degree of radial velocity anisotropy or an MM stellar
population with mass segregation (keeping isotropic velocities).
 
Nevertheless, the fit of the inner SB profile with a SM isotropic model
permitted us to determine the
best fit value for $\M/M$ regardless of velocity anisotropy,
because the presence of the latter only influences the outer region
(as happens in normal King-Michie models, see, e.g., \citealt{gunn};
\citealt{mio06}).
Thus, a grid
of SM isotropic models have been generated by sampling
the form parameters $\wbh$ and $\M/M$.
As the model profiles are expressed in dimensionless units,
they have to be scaled in both the radial and the SB dimension. 
Thus, for each model of the grid
we found the best fit values for two suitable scale
parameters, namely the ``visual'' core radius\footnote{
Here $r_\mathrm{c}$ is defined as the radius
at which the SB drops to half its value at $r_\mathrm{cu}$.
In good approximation, $r_\mathrm{c}$ coincides
with the location of the ``turn-off'' of the profile (also called `break radius'
in \citealt{noyola06}; see \citetalias{mio07} for more details).} $r_\mathrm{c}$ and the
normalisation value of the SB,
restricting the fit to data points with $r < 13\arcmin.2$.    
Since the SB measurements uncertainties are known, we minimised the 
$\chi^2$ sum weighted by the width of the error bars. 

\begin{figure}
\includegraphics[width=8.5cm]{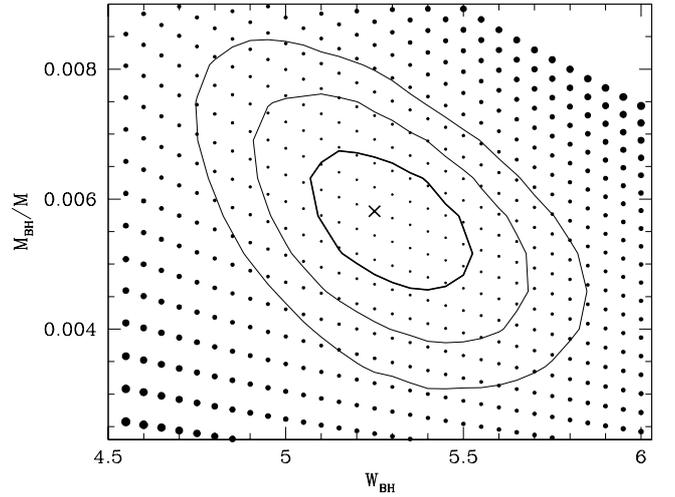}
\caption{Contours of $\chi^2$ as a function of $\wbh$ and of the $\M/M$ ratio, 
plotted for the grid of SM isotropic models (the filled dots,
with size proportional to $\chi^2$) fitted to data points
with $r<13\arcmin.2$.
The central cross marks the minimum $\chi^2$  model location ($\wbh=5.25$
and $\M/M=5.8\times 10^{-3}$)
and the isocontours for $\Delta \chi^2=2.3$, $6.17$, $11.8$ correspond to confidence regions
of $68.3$ (thick line), $95.4$, and $99.7$\%, respectively.
\label{pchi2}}
\end{figure}

The calculated $\chi^2$ values are
reported in Fig.~\ref{pchi2}, from which we deduce that
$\wbh=5.25$ and
$\M/M=5.8^{+0.9}_{-1.2} \times 10^{-3}$,
with a level of confidence (LOC) of 68.3\%.
The best fit isotropic model gives 
$r_\mathrm{c}=156 \arcsec$, which confirms both the
more recent observations by \citet{ferraro06} and the value listed
in \citet{trager95}.

\begin{figure}
\includegraphics[width=8.5cm]{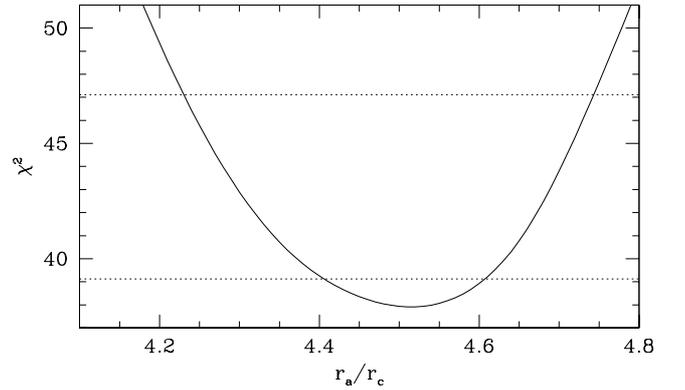}
\caption{$\chi^2$ behaviour of the best fit SM
model in the isotropic case (the cross in Fig.~\ref{pchi2}) when anisotropy is
introduced, as a function of the anisotropy radius $r_\mathrm{a}$.
In this case, all the available SB data are taken into account.
The $\chi^2$ values (for $\Delta \chi^2=1$, $9$) corresponding to an LOC of
68.3 and 99.7\% are also plotted (dotted lines). 
\label{chi2_ra}}
\end{figure}

Once $\wbh$ and $\M/M$ has been determined by fitting the inner SB profile, we 
fit the entire data set by including radial velocity anisotropy in the stellar
system outskirts (confirming what was already noted by \citealt{meylan87}),
namely outside an anisotropy radius $r_\mathrm{a}$; see, e.g., \citet{mio06}
for a description of how anisotropic velocities can be efficiently implemented in
King-Michie models.
Thus, a ``sub-grid'' of anisotropic models is generated by sampling
$r_\mathrm{a}$ in the range $[2,10]\times r_\mathrm{c}$. The resulting $\chi^2$ behaviour,
this time evaluated over all SB data,
is plotted in Fig.~\ref{chi2_ra} and leads to the estimates $r_\mathrm{a}=
(4.5\pm 0.1)\times r_\mathrm{c}=12\arcmin \pm 0\arcmin.2$ with an
LOC of 68.3\%.
From this figure, it can also be noted how the anisotropic
model when `pushed' towards the isotropic case ($r_\mathrm{a}>\!>r_\mathrm{c}$)
gives unacceptable fits (huge $\chi^2$ values).
\citet{vandeven06} find that the velocity distribution in $\omega$ Cen
is nearly isotropic inside $\sim 10\arcmin$, in agreement to our best fit
value for $r_\mathrm{a}$. On the other hand, these authors reported the presence
of a slight tangential anisotropy in the cluster outskirts which is, however,
below the uncertainty in the velocity dispersion measurements (see their figure 8).

The SM anisotropic best fit model reported in Fig.~\ref{sb}
yields a tidal radius $r_\mathrm{t}=41\arcmin.3$ and a
concentration parameter, $c=\log(r_\mathrm{t}/r_\mathrm{c})\simeq 1.2$,
substantially lower than the $1.6$ value quoted in the \citet{harris} catalogue,
but in good agreement with recent results
\citep{ferraro06,vandeven06}.

As we said at the beginning of this section, the entire observed SB profile
can also be fitted fairly well
by an MM isotropic model \citep[see also][]{meylan87} including the IMBH
and stars in the mass range $0.4$---$1.2 \msol$ distributed following
the Salpeter mass function, with central energy equipartition.
In this case, the best fit profile is for $\M/M=2.8\times 10^{-3}$ (and $\wbh=7.5$),
and it practically overlaps with that of the anisotropic case (Fig.~\ref{sb}).
Nonetheless, we discarded this MM
model because it underestimates the LOSVD in the central region, as we see in
detail in next section.

\section{Velocity dispersion profile}
\label{losvd_fit}
\begin{figure}
\includegraphics[width=8.5cm]{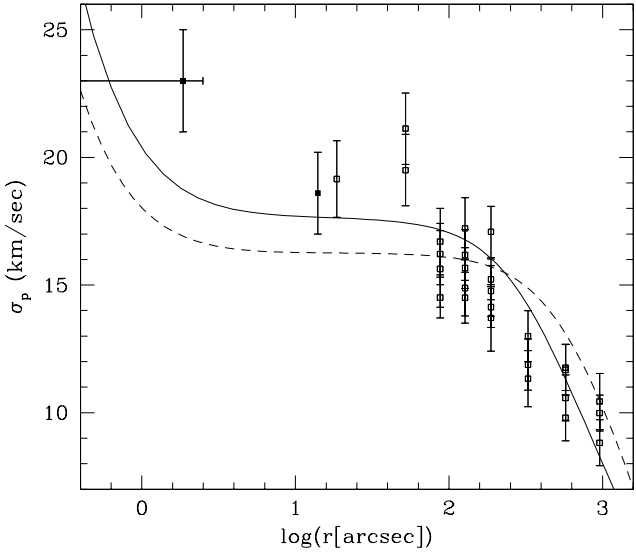}
\caption{Best fit LOSVD radial profile in the SM anisotropic (solid line)
and MM isotropic case (dashed line; in this case, it is determined by
weighting the contributions of all the stellar components according to their luminosity).
The best fit total cluster mass is $3.1$ and $4.4\times 10^6 \msol$, respectively.
Filled squares are the measurements from \citet{noyola08},
open squares come from \citet{vandeven06}. 
\label{lvd}}
\end{figure}

To provide an estimate of $\M$, we have to quantify the cluster total mass $M$ first.
This can be done by exploiting the most recent kinematic observations of $\omega$
Cen.
For this purpose, along with the two innermost points taken from the Gemini
GMOS-IFU measurements in \citet{noyola08}, we use the LOSVD data employed
by \citet[][see references therein for the discussion of the various data
sources]{vandeven06}.
These are based on various independent sets
of measurements, which in most of the radial annuli include values taken in
different apertures (see Fig.~\ref{lvd}).
A radial error bar is plotted for the innermost point to indicate 
the width of the $5\arcsec \times 5\arcsec$
GMOS-IFU field of view that was centred on the cluster nucleus to
obtain the integrated spectrum \citep{noyola08}.

Once the SB profile has been fitted, the \emph{form} of the LOSVD profile is univocally
given by the model and cannot be adapted to the observed behaviour. The best fit
can be found by adjusting only the velocity scale factor  (corresponding
to a vertical shifting of the profile). In turn, this
factor depends on the adopted cluster distance and total mass $M$.
We chose to a-priori fix the distance to $4.8$ kpc \citep[as from][]{vandeven06},
and then to find the
$M$ value that gives the best fit to the LOSVD observations.
Shown in Fig.~\ref{lvd} are two LOSVD best fit profiles: the one given by the SM
anisotropic model and the one produced by the MM isotropic one.

\begin{figure}
\includegraphics[width=8.5cm]{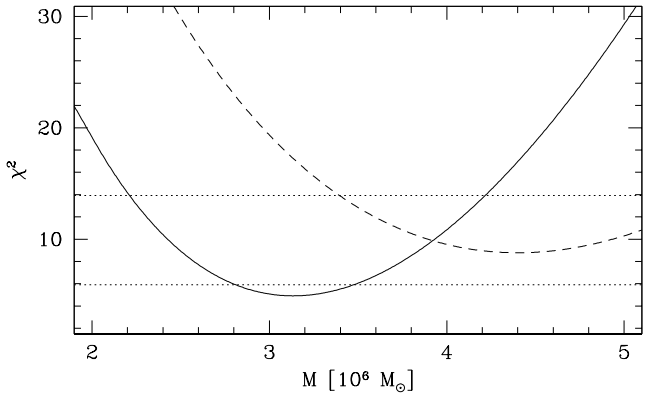}
\caption{$\chi^2$ behaviour of the LOSVD fit as a function of the total cluster mass $M$ in the
SM anisotropic (solid line) and MM isotropic (dashed) case.
The dotted lines indicate the $68.3$ and $99.7$\% LOC. 
\label{chi2_m}}
\end{figure}

It is evident (see also Fig.~\ref{chi2_m}) that the SM case yields a better fit
to LOSVD data, having $P(\chi^2 > \chi^2_\mathrm{fit}) = 84$\%,
compared with
the MM model that gives $P(\chi^2 > \chi^2_\mathrm{fit}) = 47$\% mainly because, in
the inner region ($\log r <1.8$), it exhibits too low an LOSVD.
This is naturally expected from the mass distribution in the MM
case being dominated by the lighter (and fainter) stars, which are
much less concentrated than the giants. Thus, the velocity dispersion
of the giants starts to decrease at larger radii, consequently
the best fit tends to give a lower inner LOSVD in the attempt to fit
the outer data. 
Interestingly, that the SM model better represents the
dynamical situation of this cluster suggests that mass segregation has not been
efficient in $\omega$ Cen. In this sense, it confirms the current thought that
this cluster is not completely relaxed by collisions, because of its relatively
long relaxation time (see, e.g., \citealt{meylan87,meylan95}). Various
authors, indeed, have found indications of a uniform mass-to-light ratio
\citep[see, e.g.,][]{merritt97,vandeven06}.

Considering the relatively large error bars of LOSVD measurements
in crowded regions, from Fig.~\ref{lvd} we note that the model profile
predicts that the central velocity cusp is apparently more centrally
concentrated than \citet{noyola08} observations suggest
(it starts to be evident only for $r \la 1\arcsec$), though one has to consider
the ``visual effect'' of the logarithmic scale in $r$.
In fact, a relatively large residual ($\sim 4$ km s$^{-1}$) still remains
for the innermost LOSVD data point, although its radial error bar intersects
the model profile (at $r\simeq 0\arcsec.6$).
If the \citetalias{mio07} model represents the real cluster
dynamical state well,
 this could
indicate the influence of some statistical bias affecting
this bin or too large an average radius chosen for it.
In this respect, it is also worth noting that recent and accurate proper
motion measurements reveal no significant velocity cusp at
the central region of this cluster,
though this study relies on a different dataset and kinematic centre location
and, moreover, the authors do not observe any appreciable cusp in density
\citep{anderson09}.

In \citet{noyola08} the innermost bins are fitted quite well (apart from
the measurement at $\log(r)\simeq 1.7$, see their Fig.~4).
However, it must be noticed that in these authors' model the IMBH mass best fit
value depends almost completely on the \emph{few} innermost LOSVD data points,
while it has practically no effects for $r \ga 30\arcsec$
and plays no role at all on the SB profile.
In our case, on the contrary, the behaviour of the LOSVD given by the model is strongly
dependent on the best fit parameters of the SB profile.
If our model were forced
to fit the entire LOSVD well, then the required $\sim 4\times 10^4\msol$ IMBH would
produce a much steeper SB core behaviour (along with too low a concentration), which
would be completely different from the observed one.  
The disadvantage of our parametric approach is that it is
``less general'', because it is constrained by the theoretical hypothesis
lying behind the assumption of that particular distribution function in phase-space.
  

The predicted cluster total mass is $M=(3.1\pm 0.3) \times 10^6\msol$
with an LOC of 68.3\% (Fig.~\ref{chi2_m}).
It is in marginal agreement with the dynamical estimate of $(2.5 \pm 0.3)\times 10^6\msol$
in \citet{vandeven06} -- though it would agree well at $2\sigma$ level --
while much lower than the \citet{meylan87}
$3.9\times 10^6 \msol$ value.
This author used a King-Michie MM model in which an approximated
energy equipartition was imposed \citep[see][for a discussion of this approximation]
{mio06}. This, together with the assumed presence
of very low-mass stars (down to $0.13 \msol$), can explain the higher $M$ estimate.
As far as the $M/L$ ratio is concerned, if one assumes a total $V$-band
luminosity in the ``prudential'' range $L_V=(1.0\pm 0.2)\times 10^6$ L$_\odot$
\citep[e.g.][]{seizer,meylan87,carraro},
one gets $M/L_V=3.1 \pm 0.9$, a value compatible with the accurate
$2.5 \pm 0.1$ \citet{vandeven06} estimate.
Of course, as the model predicts no mass segregation, the mass-to-light ratio
turns out to be uniform.
 
The estimate made in Sect.~\ref{surf_fit} of the ratio $\M/M$, combined with
the uncertainty on the cluster total mass, yields an IMBH mass in the range 
$1.3\times 10^4 < \M < 2.3\times 10^4 \msol$ (with a $68.3$\% LOC),
which spans about one third to a half the mass predicted by \citet{noyola08},
but is marginally compatible with the $\la 1.3\times 10^4 \msol$ \citet{anderson09b}
estimate. However, it has to be kept in mind that these two estimates rely
on different cluster centre. 
Finally, it is worth noting how our estimate range, though still incompatible, gets
closer to the
$\sim 2300 \msol$ upper limit as constrained by the $\omega$ Cen
radio continuum emission \citep{maccarone08}.
\section{Conclusions}
\label{concl}
In this paper we have presented a parametric fit of the surface brightness (SB)
profile of $\omega$ Cen (NGC 5139), made up of HST/ACS data in the central
region \citep{noyola08} and of the \citet{meylan87} normalised profile in
the outskirts.
The fit was done by using a self-consistent (spherical and non-rotating) model
that includes a central intermediate-mass black hole (IMBH).
The whole SB profile (from $r\sim 1\arcsec.6$ out to $r\sim 42\arcmin$)
can be well-fitted by the model both with the single-mass stellar distribution -- assuming
a radially anisotropic velocity distribution outside $12\arcmin$ -- and with a multimass
 model with isotropic velocity. The comparison of the
generated LOSVD with the kinematic
observations recently enriched at the very central region by Gemini
GMOS-IFU measurements \citep{noyola08}, however, allows 
this degeneracy to be resolved in favour of the single-mass case. In fact,
the multimass model yields too low an LOSVD in the central region.
This suggests that $\omega$ Cen is presently in a non mass-segregated
state, as already argued by various authors \citep[e.g.][]{meylan87,meylan95,vandeven06}.  
It is also worth noting that recent $N$-body studies show that
the presence of an IMBH in a cluster can suppress mass segregation \citep{gill08}.

From this parametric study we deduce the $68.3$\% confidence intervals
$M=(3.1\pm 0.3) \times 10^6\msol$ for the cluster total mass and 
$\M/M=5.8_{-1.2}^{+0.9} \times 10^{-3}$ for the mass ratio, 
leading to an estimate for the IMBH mass in the range $13,000 < \M < 23,000 \msol$.
This value is from about one third
to a half the $\sim 40,000\msol$ mass predicted by \citet{noyola08},
though it is compatible with the $18,000 \msol$ upper limit provided by the dynamical
analysis in \citet{anderson09b}. Note, however, that these two
published estimates are based on different cluster centre.

\section*{Acknowledgements}
The author is warmly grateful to Dr. B. Lanzoni and Dr. E. Noyola for helpful
discussions and suggestions. The paper presentation greatly benefited
from the comments of the anonymous referee.

\end{document}